# Formation of the Earth and Moon: Influence of Small Bodies

**M. Ya. Marov**[a, *] **and S. I. Ipatov**[a, **]

[a] *Vernadsky Institute of Geochemistry and Analytical Chemistry, Russian Academy of Sciences, Moscow, 119991 Russia*
*e-mail: marovmail@yandex.ru
**e-mail: siipatov@hotmail.com

**Abstract**—The paper discusses a model of the bombardment of the Earth and the Moon by small bodies when these planets were formed. It is shown that the total ice mass delivered with the bodies to the Earth from the feeding zone of the giant planets and the outer asteroid belt could have been comparable to the total mass of the Earth's oceans. Objects that initially crossed Jupiter's orbit could become Earth-crossers mainly within the first one million years. Most collisions of bodies originally located at a distance of 4 to 5 AU (astronomical units) from the Sun with the Earth occurred during the first ten million years. Some bodies from the Uranus and Neptune zones could fall onto the Earth in more than 20 million years. From their initial distances from the Sun of about 3 to 3.5 AU, some bodies could fall onto the Earth and Moon in a few billion years for the model that takes into account only the gravitational influence of the planets. The ratio of the number of bodies that collided with the Earth to the number of bodies that collided with the Moon varied mainly from 20 to 40 for planetesimals from the feeding zone of the terrestrial planets. For bodies originally located at a distance of more than 3 AU from the Sun, this ratio was mainly in the range between 16.4 and 17.4. The characteristic velocities of collisions of planetesimals from the feeding zones of the terrestrial planets with the Moon varied from 8 to 16 km/s, depending on the initial values of the semi-major axes and eccentricities of orbits of the planetesimals. The collision velocities of bodies that came from the feeding zones of Jupiter and Saturn with the Moon were mainly from 20 to 23 km/s.

## INTRODUCTION

According to current model interpretations, the Earth was formed in a high-temperature zone, and hence, in studying how life emerged on this planet, it is interesting to understand how water may have been delivered to the Earth from behind the snowline, from distances of $R > 3$ AU (Marov, 2018, 2018a). The water in the Earth's oceans and its deuterium/hydrogen (D/H) ratio could have been formed by mixing water from a number of sources with high and low D/H ratios. Some of the exogenic sources might be bodies that had migrated from the outer part of the main asteroid belt (Lunine et al., 2003, 2007; Morbidelli et al., 2000, 2012; O'Brien et al., 2014; Petit et al., 2001; Raymond et al., 2004) and planetesimals and dust that had migrated from behind the Jupiter's orbit (Levison et al., 2001; Morbidelli et al., 2000; Ipatov, 2000; Marov and Ipatov, 2001, 2005, 2020; Ipatov and Mather, 2003, 2004, 2006, 2007; Ipatov, 2010). Rubie et al. (2015) discuss the migration of planetesimals from distances of 6 to 9.5 AU from the Sun for the Grand Tack model. Ipatov (1999) suggested that close to 20% present near-Earth objects larger than 1 km in diameter may have originated from behind the Neptune's orbit. According to (Drake and Campins, 2006), a key argument against the hypothesis that water on the Earth was brought mostly from an asteroidal source is that the Os isotope composition of the primary Earth's upper mantle is closer to that of anhydrous ordinary chondrites than of hydrous carbonaceous chondrites.

One of the key and still-unresolved problem is the origin of the Moon in the Earth–Moon system. Many authors (e.g., Hartmann and Davis, 1975; Cameron and Ward, 1976; Canup and Asphaug, 2001; Canup, 2004, 2012; Cuk and Stewart, 2012; Canup et al., 2013, 2021; Cuk et al., 2016; Barr, 2016) are prone to maintain the megaimpact theory, according to which the Moon was produced by the ejection of the Earth's silicate mantle (that had already been formed by that time) when the Earth collided with a body whose size was commensurable to Mars. The megaimpact model was criticized by E.M. Galimov (Galimov et al., 2005; Galimov, 2011; Galimov and Krivtsov, 2012), who stressed that this model fails to explain the known isotope ratios of some elements (for example, similarities in the isotope composition of oxygen, iron, hydrogen, silicon, magnesium, titanium, potassium, tungsten, and chromium in the Earth and the Moon), because according to this model, the bulk of the Moon was formed of the material of the impactor but not of the proto-Earth. E.M. Galimov and his colleagues (Galimov et al., 2005; Galimov, 1995, 2008, 2011, 2013; Galimov and Krivtsov, 2012; Vasil'ev et al., 2011) have put forth a model of the origin of the Earth's and Moon's embryos by means of contraction and subsequent fragmentation of a rarefied dust condensation, whose mass was equal to that of the Earth–Moon system, in approximately 50–70 million years after the Solar System started to develop. The researchers hypothesized that such a long stability period of the original gas–dust accumulation may be explained by intense gas emanation from the surface of the particles and, perhaps, also because of ionization and radiation repulsion due to the decay of short-lived isotopes. According to this model, the Earth and the Moon were formed from a common condensation and had roughly present masses. The model proposed by E.M. Galimov is able to explain the following data: (1) the identity of the oxygen isotope composition (the $^{16}O/^{17}O/^{18}O$ isotope ratios) and the $^{53}Cr/^{52}Cr$, $^{46}Ti/^{47}Ti$, $^{182}W/^{184}W$ isotope ratios in the Earth and Moon, as opposed to other space bodies; (2) data on the U–Pb, Rb–Sr, and Hf–W isotope systems and on the depletion of siderophile elements (W, P, Co, Ni, Re, Os, Ir, and Pt) in both the Earth and the lunar mantle; (3) data on the relative enrichment of lunar basalts in major refractory elements (Ca, Al, and Ti); and (4) the ratios of the $^{57}Fe/^{54}Fe$ and $^{56}Fe/^{54}Fe$ stable isotopes measured in Luna 16, 20, and 24 lunar soil samples: these ratios are interrelated through the equation $\delta^{57}Fe = 1.48\delta^{56}Fe$, which indicates that Fe in low-Ti lunar basalts is isotopically lighter than in terrigenous basalts, but the Earth and the Moon as a whole are highly probably to possess identical Fe isotope composition, with the Fe isotopic of the Moon generally close to that of chondrites (Okabayashi et al., 2019).

The life and compression times of condensations are reportedly no longer than 1000 revolutions around the Sun (Cuzzi et al., 2008, 2010; Cuzzi and Hogan, 2012; Johansen et al., 2007, 2009, 2009a, 2011, 2012; Nesvorny et al., 2010; Lyra et al., 2008, 2009; Youdin, 2011; Youdin and Kenyon, 2013). Longer compression times can be obtained with regard to factors that impede the rapid compression of rarefied condensations. For example, depending on the optical properties of the dust and gas and on the types and concentrations of the short-lived radioactive isotopes in the condensations, the time of their compression to a density of solid bodies were no longer than a few million years (Myasnikov and Titarenko, 1989, 1990).

The origin of embryos of the Earth and Moon from a common condensation, with the masses of the embryos one to two orders of magnitude smaller than those of the present masses of these celestial bodies was discussed in (Ipatov, 2018; Marov et al., 2019). Similar to the multiple-impact model (Ringwood, 1989; Vityazev and Pechernikova, 1996; Gorkavyi, 2004, 2007; Citron et al., 2014; Rufu and Aharonson, 2015, 2017), these models analyze multiple collisions of planetesimals with the Earth's embryo and the partial fallout of the material ejected from the Earth's embryo onto that of the Moon. However, in contrast to these models, we believe (Ipatov, 2018; Marov et al., 2019) that the original Moon's embryo was formed from a common condensation from which the Earth's embryo was also formed, but not from the material ejected from the Earth's embryo. It has been demonstrated that the angular momentum of the condensation needed to form the binary system (the binary system of the Earth's and Moon's embryos) as a result of compression of the condensation could be acquired at the collision of two condensations that produced the parental condensation. This model of the origin of the Earth's and Moon's embryos is analogous to the model of the origin of the satellite systems of trans-Neptunian objects (Ipatov, 2017).

Herein we consider the probabilities and characteristic velocities of collisions with the Earth and the Moon of bodies that had migrated from different distances from the Sun and at different time. We also estimated the delivery of water and volatile components to the Earth and the Moon.

## PROBABILITIES OF COLLISIONS OF BODIES THAT HAD MIGRATED FROM DIFFERENT DISTANCES FROM THE SUN WITH THE EARTH AND MOON

In our earlier calculations of the probability $p_E$ of a collision of a body with the Earth, we simulated the evolution of the orbits of bodies that had originally formed at different distances from the Sun. When the migration of these bodies was simulated, the gravitational influence of planets was taken into account. For integration of the motion equations, we used the symplectic method of the Swift integration package (Levison and Duncan, 1994). The obtained arrays of orbital elements of the bodies were used to calculate the probabilities of collisions of bodies with the Earth and the Moon over a specified time span. In the calculation of the variant with 250 bodies, the values of $p_E$ may differ by hundreds of times for similar starting values $a_o$ of the semi-major axes and eccentricities of orbits of the bodies at $a_o \leq 10$ AU. It has been shown that the probability of a collision of a single body with the Earth may be higher than the total probability of collisions of hundreds or thousands other bodies with similar initial orbits if, during the evolution, this body has got an orbit that crossed the Earth's orbit over a time span of millions or tens of millions of years.

For planetesimals from the Jupiter and Saturn zones, the average probability of a collision of a planetesimal with the Earth was estimated at $p_E = 2 \times 10^{-6}$ (Marov and Ipatov, 2018). The mean value of the probability $p_E$ of a collision of bodies that initially had orbits of Jupiter-family comets with the Earth was higher than $4 \times 10^{-6}$ (Ipatov and Mather, 2003, 2004, 2006, 2007). The $p_E$ values were, on average, smaller for higher initial values of the semi-major axes $a_o$ of the orbits (Marov and Ipatov, 2020; Ipatov, 2020). For the Uranus and Neptune feeding zone, $p_E$ was close to $10^{-6}$. At $3 \leq a_o \leq 4$ AU in some calculation variants, $p_E > 10^{-4}$, although $p_E = 2 \times 10^{-6}$ for other variants.

At $p_E = 2 \times 10^{-6}$ and with the assumption that the total mass of planetesimals in the Jupiter and Saturn feeding zone was 100 Earth's masses (Ipatov, 1993, 2000), we found out that the total mass of planetesimals that have fallen onto the Earth was $2 \times 10^{-4} m_E$ (where $m_E$ is the Earth's mass). About the same amount of bodies could migrate to the Earth from the outer asteroid belt and from behind the Saturn's orbit. If the ice made up half of the mass of this material, then the total ice mass delivered to the Earth from behind the snowline was equal to the total mass of the Earth's oceans ($\sim 2 \times 10^{-4} m_E$). Probably, the ice fraction in the planetesimals was a little lower than a half (for example, it was close to one-third). Indeed, estimated ice amounts in comets are no higher than 33% (Greenberg, 1998; Davidsson et al., 2016; Fulle et al., 2017). However, some authors believe that the primary planetesimals may have contained more ice than it is now found in comets. The conclusion that such an amount of water could have been brought to the Earth from the feeding zone of the giant planets was also arrived to in (Ipatov, 1995, 2000, 2001; Marov and Ipatov, 2001). The $p_E$ value for planetesimals from the Uranus and Neptune feeding zone was estimated at $1.2 \times 10^{-6}$ (Ipatov, 2000). In these estimates, it was assumed that the fraction of planetesimals that have reached the Earth's orbit was 0.1, the characteristic time during which a planetesimal that had reached the Earth's orbit crossed it was $5 \times 10^3$ years, and the probability of a collision of such a planetesimal (that had acquired the orbit that crossed the Earth's orbit) with the Earth per year was obtained to be $2.5 \times 10^{-9}$. In the estimates presented above for water delivery onto the Earth and in (Ipatov, 1993, 2000), the total mass of bodies behind the Jupiter's orbit was estimated to be about $200 \times$ Earth's mass. The mass of the planetesimal disk, up to $200 m_E$, was also discussed in (Hahn and Malhotra, 1999). It has been found out (Canup and Pierazzo, 2006) that a collision of a planetesimal with the Earth at a velocity higher than $1.4 \times$ parabolic velocity and an angle greater than $30°$ shall result in losses of more than 50% water of the impactor. Because of this, the amount of water accumulated by the terrestrial planets may have been lower than the water amount delivered to these celestial bodies.

Bodies from different distances from the Sun may have reached the Earth at different times. Former Jupiter-crossing bodies could become Earth-crossers within the first million years. Bodies that contained water and volatiles migrated toward the terrestrial planets also from behind the Saturn's orbit and from the outer asteroid belt. The migration time of bodies from the feeding zone of Uranus and Neptune depends on when large embryos of these planets appeared in this zone. The main changes in elements of the orbits of the embryos of the giant planets (Ipatov, 1993, 2000) took place within no more than 10 million years, although some bodies could fall onto these planets in time spans of about billions of years. Our calculations of the migration of the bodies show that, at the present orbits and masses of the planets, some bodies from the Uranus and Neptune zone may have fallen onto the Earth in more than 20 million years. At an initial value $a_o$ of the semi-major axis of the body's orbit within the range of 4 to 5 AU, most bodies fell onto the Earth within 10 million years. At $a_o$ close to 3 AU and at small eccentricities of initial orbits, the first bodies could begin reaching the Earth's orbit in more than one billion years. Some bodies with $3 \leq a_o \leq 3.7$ AU could fall onto the Earth in billions of years, and such bodies could be involved in the late heavy bombardment (LHB) of the Earth and the Moon.

In the calculation variants discussed herein, the ratio $r_{EM}$ of the number of bodies that have fallen onto the Earth to the number of bodies fallen onto the Moon was smaller than the mass ratio of these celestial objects. Calculations of the migration of planetesimals in the feeding zone of the terrestrial planets (Ipatov, 2019) may lead to the conclusion that the ratio $r_{EM}$ of the number of planetesimals colliding with the Earth to that colliding with the Moon mostly varied from 20 to 40. The reason for the scatter of this ratio is that the considered planetesimals have different initial values of the semi-major axes and eccentricities of the orbits. For planetesimals with orbital semi-major axes $0.9 \leq a_o \leq 1.1$ AU and with small eccentricities, the $r_{EM}$ ratio was mostly within the range of 30 to 40. Planetesimals initially more distant from the Earth's orbit came to it from more eccentric orbits, and their $r_{EM}$ ratio was smaller than that for more close orbits with small eccentricities.

In considering the model in which the masses of the embryos of the terrestrial planets and the Moon were tenfold smaller than the present masses of these celestial bodies, the $r_{EM01}$ ratio of the number of planetesimals that collided with the Earth's and Moon's embryos varied mostly from 23 to 26 for planetesimals with $0.7 \leq a_o \leq 1.1$ AU and from 17 to 20 at $0.5 \leq a_o \leq 0.7$ AU and $1.1 \leq a_o \leq 1.3$ AU. The ratio of the number of planetesimals that collided with the Earth's and Moon's embryos reached 54 (Ipatov, 2019) at the migration of the planetesimals from $0.3 \leq a_o \leq 0.5$ AU and for masses of the embryos equal to 0.3 of the current masses of the celestial objects.

**Table 1.** Relative entry velocity $v_{relE}$ of a body into the Earth's sphere of action and the collision velocities $v_{colE}$ and $v_{colM}$ of the bodies with the Earth and the Moon for some $r_{EM}$ values of the ratio of the number of planetesimals that collided with the Earth and Moon

| $r_{EM}$ | 14.56 | 16.4 | 16.6 | 17 | 17.4 | 17.89 | 20 | 30 | 40 |
|---|---|---|---|---|---|---|---|---|---|
| $v_{relE}$, km/s | 37.82 | 23.14 | 22.38 | 21.07 | 19.96 | 18.81 | 15.42 | 9.56 | 7.35 |
| $v_{colM}$, km/s | 37.89 | 23.26 | 22.51 | 21.20 | 20.10 | 18.96 | 15.60 | 9.85 | 7.73 |
| $v_{colE}$, km/s | 39.44 | 25.70 | 25.02 | 23.85 | 22.88 | 21.88 | 19.05 | 14.71 | 13.38 |

Our recent calculations of the migration of the bodies from $3 \leq a_o \leq 5$ AU may lead to the conclusions that $16.4 \leq r_{EM} \leq 17.4$ in approximately 80% of the calculation variants. In other calculation variants, the $r_{EM}$ ratio could be 14.56 and 17.89. According to (Marov and Ipatov, 2018), the $r_{EM}$ ratio of the number of planetesimals that collided with the Earth and the Moon averaged over 2000–2500 bodies varied mostly from 16.47 to 16.72 in each calculation set for planetesimals from the Jupiter and Saturn feeding zone (with $5 \leq a_o \leq 12$ AU).

If the gravitational influence of celestial bodies is not taken into account, then the probability of a collision of a small body with an object of mass $m$ is proportional to the squared radius of this object and is also proportional to $m^{2/3}$ at their equal density. The actual ratio of the Earth's and Moon's radii is 3.667, and the squared ratio is 13.45. Because the Earth's gravitational influence is greater than that of the Moon, the obtained $r_{EM}$ value is greater than 13.45. The squared effective radius $r_{ef}$ of a celestial body of radius $r$ is $r_{ef}^2 = r^2[1+(v_{par}/v_{rel})^2]$,  (1) where $v_{par}$ is the parabolic velocity at the surface of this celestial body, and $v_{rel}$ is the relative velocity of the small body when it enters the sphere of action of the celestial object (the general formula is valid at an infinite relative velocity).

## CHARACTERISTIC VELOCITIES OF COLLISIONS OF BODIES WITH THE MOON AND THE EARTH

In the calculations (Nesvorný et al., 2017), the average velocities of collisions of asteroids whose initial orbital semi-major axes were within the range of 1.6 to 3.3 AU with the Earth varied from 21 to 23.5 km/s.

Based on the $r_{EM}$ ratio of the number of planetesimals colliding with the Earth and the Moon, it is possible to estimate the characteristic velocities $v_{relE}$ (relative to the Earth) of planetesimals when they enter the Earth's sphere of action. Taking into account that $r_{EM}$ is the ratio of the squared Earth's and Moon's effective radii and using Eq. (1), we obtain for the $v_{relE}/v_{parE}$ ratio of the relative velocity of a body when it enters the sphere of action of the Earth to the parabolic velocity $v_{parE} = 11.186$ km/s at the Earth's surface

$$(v_{relE}/v_{parE})^2 = [r_{EM}(v_{parM}/v_{parE})^2(r_M/r_E)^2 - 1]/[1 - r_{EM}(r_M/r_E)^2], \quad (2)$$

where $r_M$ and $r_E$ are the Earth's and Moon's radii, and $v_{parM}$ and $v_{parE}$ are the parabolic velocities at the Earth's and Moon's surface, respectively. Table 1 lists $v_{relE}$, $v_{colM}$, and $v_{colE}$ values for some $r_{EM}$ values obtained with regard to the relations $v_{colE}^2 = v_{relE}^2 + v_{parE}^2$ and $v_{colM}^2 = v_{relM}^2 + v_{parM}^2$, for the velocities $v_{colE}$ and $v_{colM}$ of collisions of bodies with the Earth and the Moon and assuming $v_{relE} = v_{relM}$.

Taking into account the $r_{EM}$ values presented above, the following conclusions can be drawn concerning the characteristic collision velocities of bodies with the Earth and the Moon of present masses for some cases. For planetesimals with semi-major axes of their initial orbits $0.9 \leq a_o \leq 1.1$ AU and with relatively small initial eccentricities (at $30 \leq r_{EM} \leq 40$), the characteristic velocities of their collisions with the Earth lied mostly within the range of 13 to 15 km/s, and the collision velocities with the Moon were 8 to 10 km/s. For planetesimals coming from other parts of the feeding zone of the terrestrial planets (at $20 \leq r_{EM} \leq 40$), the scatter of the characteristic velocities of collisions of the planetesimals with the Earth was mostly within the range of 13 to 19 km/s, and that with the Moon was 8 to 16 km/s. For most bodies with semi-major axes of their initial orbits of 3 to 5 AU (at $16.4 \leq r_{EM} \leq 17.4$), the analogous scatter of the collision velocities was 23 to 26 km/s for the Earth and 20 to 23 km/s for the Moon. However, the range of velocities of all bodies from this zone (at $14.56 \leq r_{EM} \leq 17.89$) was wider: 22 to 39 km/s for the Earth and 19 to 38 km/s for the Moon.

At smaller masses of the Earth's and Moon's embryos, the collision velocities were smaller. Ipatov (2019) presents the results of calculations of the migration of planetesimals whose initial values of semi-major axes were within the range of $a_{0min}$ to $a_{0min}+0.2$ AU (the $a_{0min}$ values were varied from 0.3 to 1.3 AU), and whose initial orbital eccentricities were 0.05 and the initial inclinations were (in radians) 0.025. The probabilities of collisions of planetesimals with the embryos of the terrestrial planets and the Moon whose masses were $k_m$ of the present masses of the planets and the Moon ($k_m$ was assumed to be 0.1, 0.3, or 1) were calculated. For a time span $T$ of 1, 5, or 20 million years and $k_m$ = 0.1, the ratio $r_{EM01}$ of the number of planetesimals that collided with the Earth's and Moon's embryos is presented in Table 2. In fact, the masses of the embryos could have significantly changed over a time span of no longer than one million years. Data on a span of 20 million years are presented for a larger statistics. Ipatov (2019) arrived at the conclusion that the Earth and Moon have acquired more than half of their masses over a time period of no longer than five million years. Using a formula analogous to (2), we have calculated the $v_{relE01}$ values (Table 2) of the relative entry velocities of planetesimals into the action sphere of the Earth's embryo. Table 2 also shows the values $v_{colE01}$ and $v_{colM01}$ of velocities of collisions of planetesimals with the Earth's and Moon's embryos for some values of the $r_{EM01}$ ratio. The data for $T$ = 20 million years correspond to a greater increase in the average orbital eccentricities of the planetesimals and to generally majorizing estimates of the collision velocities. However, the scatter of velocities for the considered variants with different $T$ values discussed herein was relatively small. As seen in Table 2, the characteristic collision velocities of planetesimals that initially located relatively close to the orbit of the Earth's embryo, with the Earth's and Moon's embryos, whose masses were tenfold smaller than the present masses of these celestial objects, were generally within the range of 7 to 8 km/s for the Earth's embryo and 5 to 6 km/s for the Moon's embryo. For planetesimals coming from more distant (relative to the Earth's orbit) regions of the feeding zone of the terrestrial planets, the characteristic velocities ranged from 9 to 11 km/s for collisions with the Earth's embryo and from 7 to 10 km/s for collisions with the Moon's embryo.

**Table 2.** Relative entry velocity $v_{relE01}$ of planetesimals into the sphere of action of the Earth's embryo and the collision velocities $v_{colE01}$ and $v_{colM01}$ of planetesimals with the Earth's and Moon's embryos for some values of $r_{EM01}$ ratios of the number of planetesimals that collided with the Earth's and Moon's embryos. The masses of the embryos were ten times smaller than the present Earth's and Moon's masses. The initial values of the semi-major axes of the orbits of the planetesimals that corresponded to $r_{EM01}$ were within the range of $a_{0min}$ to $a_{0min}$ + 0.2 AU, $T$ is the considered time span in million years

| $a_{0min}$, AU | 0.5 | 0.5 | 0.7 | 0.7 | 0.9 | 0.9 | 1.1 | 1.1 |
|---|---|---|---|---|---|---|---|---|
| $T$, million years | 5 | 20 | 1 | 20 | 1 | 20 | 5 | 20 |
| $r_{EM01}$ | 19.8 | 18.1 | 25.6 | 23.2 | 24.0 | 23.5 | 17.3 | 17.1 |
| $v_{relE01}$, km/s | 7.30 | 8.55 | 5.22 | 5.85 | 5.62 | 5.57 | 9.41 | 9.67 |
| $v_{colE01}$, km/s | 8.96 | 10.01 | 7.36 | 7.82 | 7.65 | 7.61 | 10.75 | 10.97 |
| $v_{colM01}$, km/s | 7.38 | 8.62 | 5.24 | 5.96 | 5.73 | 5.67 | 9.48 | 9.73 |

The values presented in Tables 1 and 2 for the characteristic collision velocities of the bodies with the Moon and the Moon's embryo are several times higher than, respectively, the parabolic velocities on the surface of the Moon ($v_{parM}$ = 2.38 km/s) and on the surface of the Moon's embryo whose mass was ten times smaller than the Moon's mass ($v_{parM0}$ = 1.1 km/s). For example, $v_{colM}/v_{parM}$=8.8 at $v_{colM}$=21 km/s and $v_{colM01}/v_{parM01}$=5.2 at $v_{colM01}$=5.7 km/s.

## CONCLUSIONS

This paper discusses the delivery of the material of small bodies/planetesimals to the Earth and the Moon in the course of their growth. It is demonstrated that the total mass of water ice brought to the Earth from the feeding zone of the giant planets and from the outer asteroid belt may have been comparable to the total mass of the Earth's oceans. Planetesimals that initially crossed the Jupiter's orbit could reach the Earth's orbit mostly within the first million years after the origin of the Solar System. Bodies that had initially been located at distances from 4 to 5 AU from the Sun mostly fell onto the Earth within the first ten million years, whereas bodies from the Uranus and Neptune zone may have fallen onto the Earth within more than twenty million years. Some bodies from 3 to 3.5 AU from the Sun may have fallen onto the Earth in a few billion years (if only the gravitational influence of planets is taken into account).

The ratio of the number of bodies that collided with the Earth and Moon generally varied from 20 to 40 for planetesimals from the feeding zone of the terrestrial planets. For bodies that migrated from distances >3 AU from the Sun, this ratio was mostly within the range of 16.4 and 17.4. The characteristic collision velocities of planetesimals from the feeding zones of the terrestrial planets with the Moon varied from 8 to 16 km/s, depending on the initial distances of these planetesimals from the Sun and on their eccentricities. The collision velocities of bodies that migrated from Jupiter's and Saturn's feeding zones with the Moon were generally within the range of 20 to 23 km/s. The characteristic collision velocities of planetesimals that were initially located at 0.7 to 1.1 AU from the Sun with the Earth's and Moon's embryos, whose masses were ten times smaller than the present masses of these celestial objects, were mostly within the range of 7 to 8 km/s for the Earth's embryo and 5 to 6 km/s for the Moon's embryo. For planetesimals that came from regions more distant from the Earth's orbit, in the feeding zone of the terrestrial planets, the characteristic velocities were 9 to 11 km/s for collisions with the Earth's embryo and 7 to 10 km/s for collisions with the Moon's embryo.

## FUNDING

The studies of falls of bodies onto the Earth were carried out under government-financed research project 0137-2019-0004 for the Vernadsky Institute of Geochemistry and Analytical Chemistry, Russian Academy of Sciences. The studies of falls of bodies onto the Moon were supported by the Russian Science Foundation, project 21-17-00120, https://rscf.ru/project/21-17-00120/.